\documentclass[10pt,aps,prl,twocolumn,showpacs,superscriptaddress,longbibliography]{revtex4-1}
\usepackage{lmodern}
\usepackage{amsmath}
\usepackage{graphicx}
\usepackage{color}
\usepackage{natbib}
\usepackage{physics}

\usepackage[colorlinks,bookmarks=false,citecolor=blue,linkcolor=blue,urlcolor=blue,hyperfootnotes=true]{hyperref}

\begin{document}
\title{Role of the quasi-particles in an electric circuit with Josephson junctions}

\author{Beno\^it Rossignol}
\author{Thomas Kloss}
\author{Xavier Waintal}
\affiliation{Univ.\ Grenoble Alpes, CEA, INAC-Pheliqs, 38000 Grenoble, France}

\date{\today}

\begin{abstract}
While Josephson junctions can be viewed as highly non-linear impedances for superconducting quantum technologies, they also possess internal dynamics that may strongly affect their behavior. Here, we construct a computational framework that includes a microscopic description of the junction (full fledged treatment of both the superconducting condensate and the quasi-particles) in presence of a surrounding electrical circuit. Our approach generalizes the standard Resistor-Capacitor-Josephson model (RCJ) to arbitrary junctions (including e.g. multi-terminal geometries and/or junctions that embed topological or magnetic elements) and arbitrary electric circuits
treated at the classical level. By treating the superconducting condensate and quasi-particles on equal footings, we capture non-equilibrium phenomena such as Multiple Andreev Reflection.
We show that the interplay between the quasi-particle dynamics and the electrical environment leads to the emergence of new phenomena. In a RC circuit connected to single channel Josephson junction, we find out-of-equilibrium current-phase relations that are strongly distorted with respect to the (almost sinusoidal) equilibrium one, revealing the presence of high harmonic AC Josephson effect.
In an RLC circuit connected to a junction, we find that the shape of the resonance is strongly modified by the quasi-particle dynamics: close to resonance, the current can be {\it smaller} than without the resonator. Our approach provides a route for the quantitative modeling of superconducting based circuits.
\end{abstract}
\maketitle

There is currently a huge effort around the world - both within academia and major industrial partners - to promote the superconducting transmon quantum bit \cite{Koch2007, Barends2013, Paik2011} from a laboratory object to a viable technology for building a quantum computer. The central element of this approach is a weak normal link between two pieces of superconductors, the Josephson junction. Although tunneling junctions with an insulating (oxide) barrier are the most mature elements, other types of junctions such as atomic contacts \cite{Goffman2000} (with very few propagating channels),  semiconducting nanowires \cite{Mourik2012} (with high spin-orbit suitable for stabilizing  Majorana bound states) superconducting-ferromagnetic-superconducting \cite{Izyumov2002, Buzdin2005} (with anomalous current phase relations) or multiterminal devices \cite{Riwar2016} could provide new functionalities to the superconducting toolbox. While the theoretical description of these objects is rather well understood \cite{Golubov2004}, many relevant regimes lie outside of what may be treated analytically and the development of numerical methods is important. In fact, the complexity of the circuits that are being created is increasing very rapidly and building predictive numerical tools is a key element for the success of any quantum technology.

\begin{figure}
\centering
\includegraphics[scale=0.6]{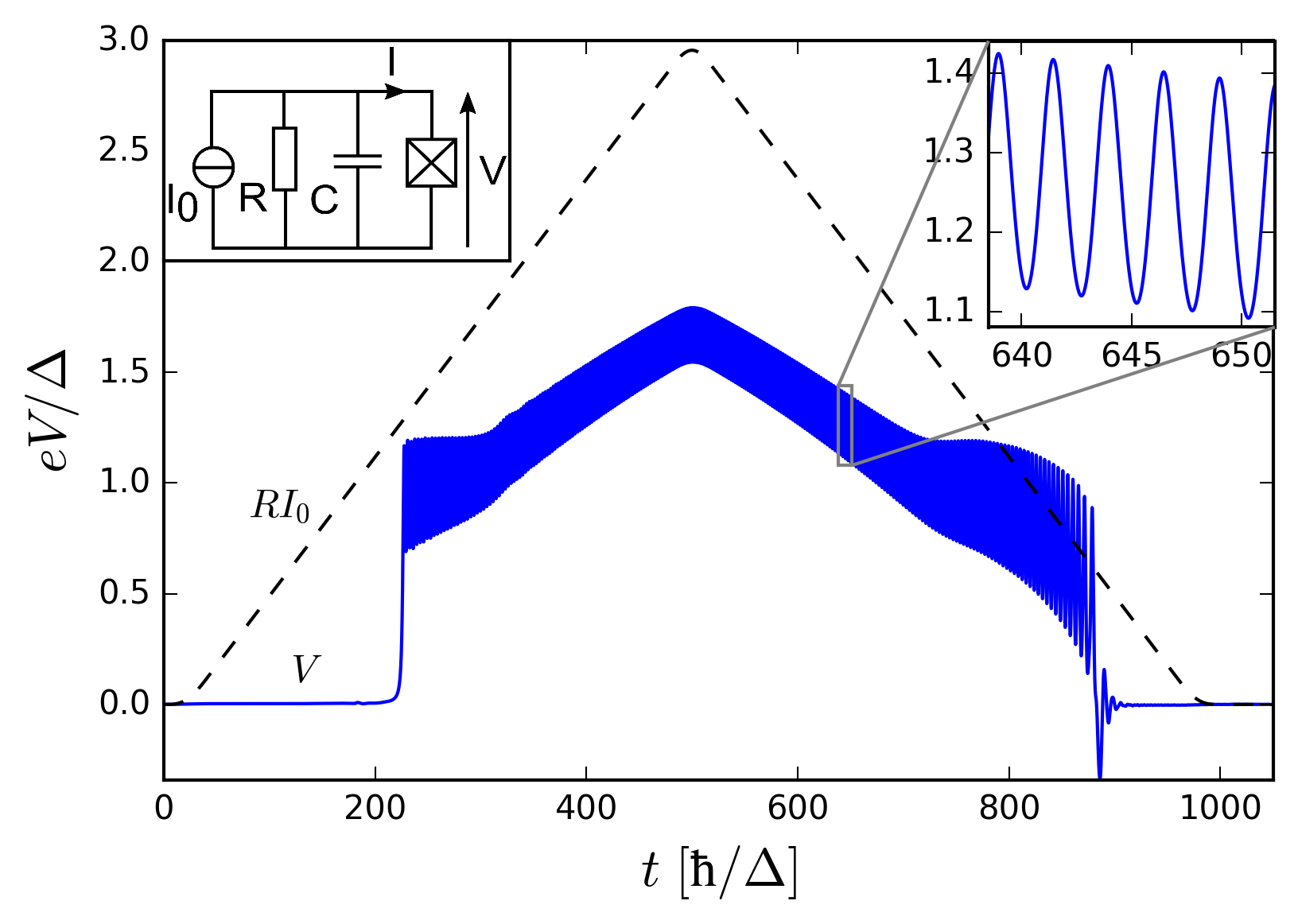}
\vspace{-3ex}
\caption{Upper left inset the simulated circuit, an RC biased Josephson junction. Main panel result of the RC-BdG simulation. Dashed line: voltage $R I_0$ applied by the generator versus time $t$. $I_0$ is raised and decreased slowly to keep the system quasi-adiabatic. Blue line: voltage $V(t)$ measured across the junction. Upper right inset: zoom of the main curve revealing the oscillations due to the AC Josephson effect.}
\label{fig1}
\end{figure}

Two very successful complementary viewpoints are commonly used to describe Josephson junctions circuits. The first one is the RCJ model \cite{Likharev1979, Stewart1968} (Resistor-Capacitor-Josephson) that views the Josephson junction as a highly non-linear impedance embedded in an electric circuit. In this model, one considers a classical circuit such as the ones shown in the insets of Fig. \ref{fig1} or Fig. \ref{fig3} with resistances $V = R I$,
capacitances $I = C \partial_t V$, inductances $V = L\partial_t I$ or other classical elements. The Josephson junction is described by its current-phase relation
$I = I_c \sin \varphi$ and the Josephson relation $\partial_t \varphi = (2e/\hbar) V$.
Such a simple model is surprisingly powerful. It captures the hysteresis loops of the $I-V$ curves. Its simple extension, where one adds a Langevin stochastic term to account for finite temperature, accurately describes the noise properties found experimentally including the probability for the junction to switch from the superconducting branch \cite{BenJacob1982}. It has also been successfully used for more elaborate circuits that include resonators \cite{Cassidy2017}.
Its quantum extension provides the model used to design the various sorts of superconducting qubits \cite{Xiang2013} and has been shown to describe very accurately a large corpus of experimental data \cite{Devoret1990}. Yet, the model fails dramatically in some very simple limits. For instance, at large voltages, it does not properly reproduce the Ohmic behavior of the circuit, since the latter involves the excitation spectrum of the junction which is not accounted for in the current-phase relation. More importantly, it does not account for some important out-of-equilibrium phenomena 
such as Multiple Andreev Reflexion \cite{Klapwijk1982,Averin1995} (MAR) processes.

The second model uses a microscopic mean-field description of the junction through the (time-dependent) Bogoliubov-De-Gennes (BdG) equation. BdG models capture most of the salient features of these junctions including those which contain exotic (e.g. topological or magnetic) materials. It naturally describes MAR \cite{Averin1995}, the interplay with microwaves \cite{Cuevas2002}, ac Josephson effects and emergent topological effects in multi-terminal geometries \cite{Houzet2013}. Until recently, however, the direct numerical integration of BdG equations has been very limited due to the intrinsic computational complexity \cite{Weston2016} and did not include the electromagnetic environment of the junction. 

The present letter builds on recent advances made in time-dependent computational transport \cite{Kloss2018} to construct a numerical method that merges the RCJ model with the BdG equation, thereby providing a fully self-consistent treatment of the Josephson junction and its electromagnetic environment at the BdG level (hereafter called RC-BdG model). The method has an arbitrary precision and is scalable to hundreds of thousands of orbitals, paving the way to the simulations of complex superconducting circuits. It applies to arbitrary BdG equations and classical electromagnetic environments.

{\it Problem formulation.} We model our circuits in two parts. First, the junction itself is described with a microscopic BdG Hamiltonian $\hat H(\varphi,t)$ that depends explicitly on time $t$ (through e.g. a capacitive gate) and on the phase difference $\varphi(t)$ between the two superconductors (which extends to a vector when more than two superconductors are involved). Note that due to the AC Josephson effect, the problem is intrinsically time-dependent even in the absence of time-dependent perturbations. Integrating the BdG equation provides the density matrix $\hat\rho (t)$ from which one can compute the current $I(t)$ that flows through the system. Below, we restrict ourself to the average current, but its quantum fluctuations are also accessible through our formalism \cite{Gaury2016}. The second part of the model describes the classical circuit, or the electromagnetic environment, that surrounds the junction. The classical equations that
describe these circuits take the form of a differential equation for $\varphi$. More complex circuits are described in a similar way with more degrees of freedom describing the classical circuit. The set of equations reads,
\begin{subequations}
\begin{align}
&i\hbar \partial_t \hat \rho = [\hat H(\varphi, t),\hat \rho] ,
   \label{eqa}\\
&I(t) = {\rm Tr } (\hat I \hat \rho )               ,           \label{eqb} \\
& \frac{d^2 \varphi}{d t^2} = F\left(\varphi, \frac{d \varphi}{d t}, I(t)\right)\label{eqc},
\end{align}
\end{subequations}
where the function $F\left(\varphi, d \varphi/d t, I(t)\right)$ describes the dynamics of the classical circuit (RC or RLC equation in the examples below) with $I(t)$ as an external source term.
We numerically solve the BdG equation within the Keldysh formalism using the approach developed in \cite{Gaury2014, Weston2016b} where the problem is unfolded onto a set of
Schr\"odinger equations for scattering wave functions. An efficient algorithm has been constructed in \cite{Weston2016} to integrate the corresponding equations. The corresponding software, ``t-Kwant'' relies on the Kwant package \cite{Groth2014} and will be released as open source in a near future. Eq.\eqref{eqa} alone amounts to solving a few hundred time-dependent Schr\"odinger equations (the actual number depending on the required energy resolution). The self-consistent condition Eq.\eqref{eqb} and \eqref{eqc} makes the problem significantly more challenging since it creates non-linear couplings between these Schr\"odinger equations. Following \cite{Kloss2018}, we address this non-linear coupling by taking advantage of the separation of time scales in the problem between the microscopic time scales of the BdG equation (which imposes discretized  time steps of lengths much smaller than $\hbar/E_F$ with $E_F$ the  Fermi energy) and the evolution of the electromagnetic variables $I(t)$ and $\varphi(t)$ that takes place on much slower time scales (typically GHz frequencies as compared with PHz for the Fermi energy in actual devices). Hence, we use a doubly adaptive predictor-corrector approach for $\varphi (t)$ as explained in \cite{Kloss2018}: Eq.\eqref{eqa} is integrated with a ``predicted'' function $\varphi (t)$ and  Eq.\eqref{eqb} and \eqref{eqc} are used on the larger time scale to construct this prediction.
A straightforward time adaptive fourth order Runge-Kutta \cite{Hairer1993} is used for the integration of Eq.\eqref{eqc}. 
We used the algorithm of \cite{Istas2018} to calculate the (Andreev) bound states of the model with precision.
The method is general to any Hamiltonian $\hat H$ that is quadratic in creation/destruction operators. Hence,
it may handle electron-electron interaction effects at the mean field or Random Phase Approximation level\cite{Kloss2018} but does not capture correlation effects. Static or dynamic disorder may be added directly\cite{Gaury2014}. The full code used for generating the data of this article can be found in \cite{code}

To be specific, we now turn to a particular BdG Hamiltonian that describes a single channel junction. 
 The BdG Hamiltonian describes a one dimensional systems with the two superconductor corresponding to $x<0$ and $x>0$ while the normal region is formed by a single site at $x=0$ placing the system in the short junction limit,
\begin{align}
\hat H &= \sum_{\substack{x=-\infty \\ \sigma =\uparrow,\downarrow}}^{+\infty}
 e^{i\varphi(t)\delta_{x,-1}} \hat c^\dagger_{x\sigma}\hat c_{x+1,\sigma}   +   (U\delta_{x,0} - E_F) \hat c^\dagger_{x\sigma}\hat c_{x\sigma}   \nonumber \\
&+\sum_{x=-\infty}^{+\infty} \Delta (1-\delta_{x,0}) \hat c^\dagger_{x\uparrow}\hat c^\dagger_{x\downarrow} + h.c.  
\end{align}
Here $\varphi(t)=(e/\hbar)\int^t V(t')dt'$ where $V(t)$ is the voltage difference across the junction, $\Delta$ is the superconducting gap inside the superconductors, $U$ is a
potential barrier used to tune the transmission probability $D$ of the junction. In the calculations below we used $E_F=2$, $\Delta=0.1$ (which will be used as our unit of energy),
and $U=2$ which corresponds to a junction with intermediate transmission $D=0.5$. For this value, the equilibrium  current-phase relation has small deviations with respect to a sinusoidal shape but the $I-V$ characteristics of the isolated junction exhibits distinct cusps at voltages 
$eV=\Delta/n$, $n\in\{1,2,3\dots\}$ (MAR) \cite{Averin1995}. The precise relation $I(\varphi) = I_c\sin(\varphi)$ is recovered in the tunneling regime $D\ll 1$ and $eV < \Delta$ with $I_c = 2e \Delta D/h$.

{\it Results for the RC-BdG model.} The first electromagnetic environment we consider is a simple RC circuit as sketched in Fig. \ref{fig1}. The capacitance $C$ typically accounts for the electron-electron interaction in the junction itself while the resistance $R$ accounts for the finite residual resistance in the whole circuit. This $RC$  circuit is the minimum electromagnetic environment that must be considered. The RCJ model for this circuit (where the BdG equation is replaced by the current-phase relation)
reads,
\begin{equation}
\frac{d^2\varphi}{dt^2}+\frac{1}{Q} \frac{d\varphi}{dt}+\sin(\varphi)=\frac{I_0}{I_c},
\end{equation}
where the time $t$ has been rescaled as $t\rightarrow \omega_0 t$. $\omega_0 = \sqrt{\hbar I_c/(2e C)}$ is the intrinsic
frequency of the circuit for small oscillating amplitudes and $Q = RC \omega_0$ is the corresponding quality factor. The physics of this model is well understood \cite{Stewart1968}: for $I_0 < I_c$ all the current passes through the junction (super-current branch) while for $I_0 > I_c$ the equilibrium solution is unstable and a voltage develops across the junction. Interestingly, this model is hysteretic for underdamped circuits $Q > 1$ and a dynamical solution with $\dot\varphi\neq 0$ exists for some values of $I_0 < I_c$. At high bias current $I_0\gg I_c$, most of the current is dissipated by the resistor $R$ and the RCJ model predicts $I_0 = R \bar V$ (where $\bar V$ is the average voltage difference seen by the junction. This prediction misses an important contribution from the junction, its intrinsic resistance $R_J = h/(2e^2 D)$ in the normal state. Indeed, at large bias current, one expects $I_0 = (1/R+1/R_J)\bar V$

\begin{figure}
\centering
\includegraphics[scale=0.6]{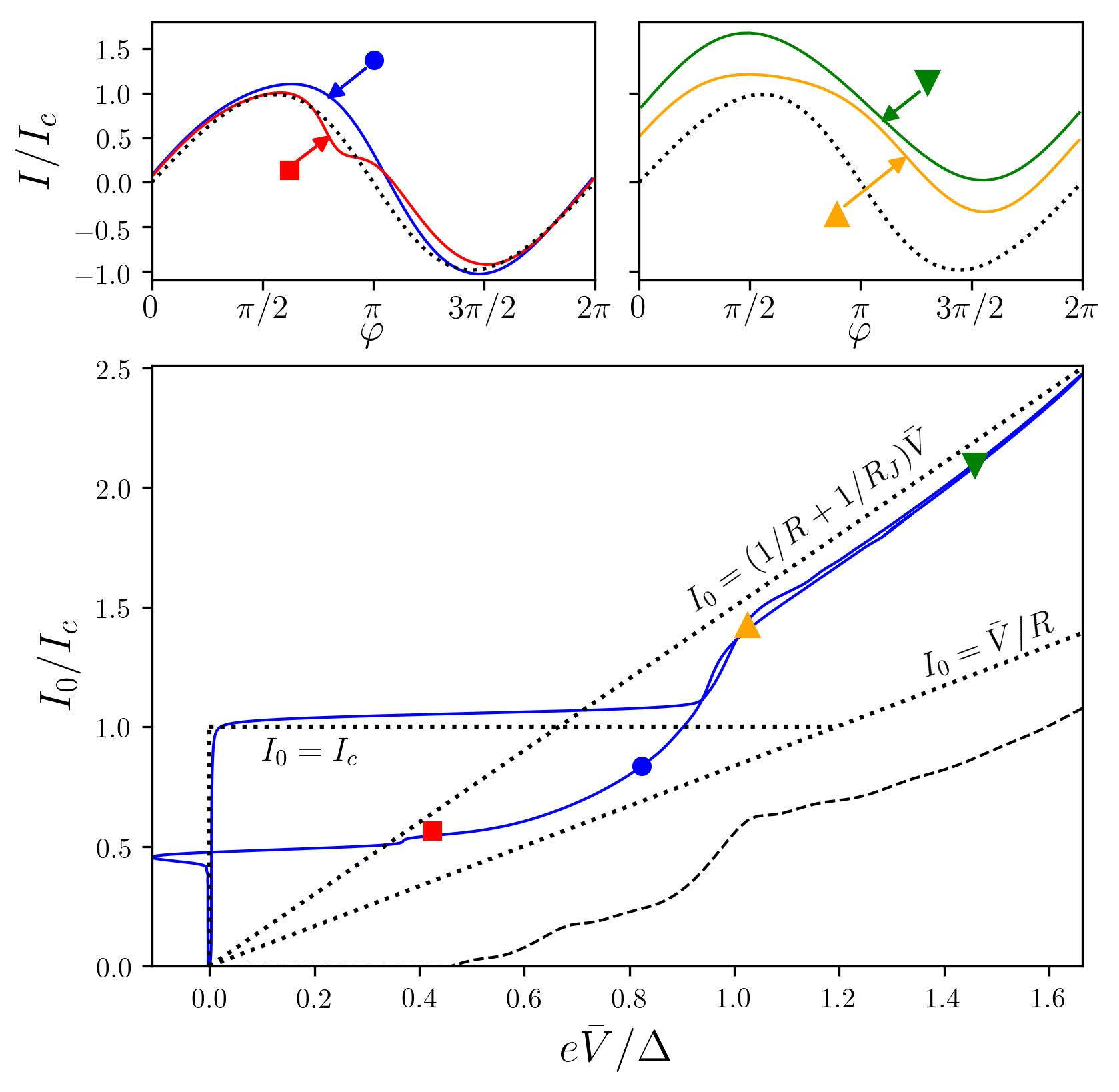}
\vspace{-3ex}
\caption{RC-BdG model. Bottom panel: bias current $I_0$ versus the average voltage across the junction $\bar V$ for an underdamped oscillator $Q\approx 1.7$. Dotted lines: various asymptotes of the RCJ model, see text. Dashed line:
pure BdG model without the environment. Upper panels: out-of-equilibrium current-phase relations at 4 different points of the
$I_0-\bar V$ curve. The dotted line corresponds to the equilibrium current-phase relation of the pure junction.}
\label{fig2}
\end{figure}

\begin{figure}
\centering
\includegraphics[scale=0.5]{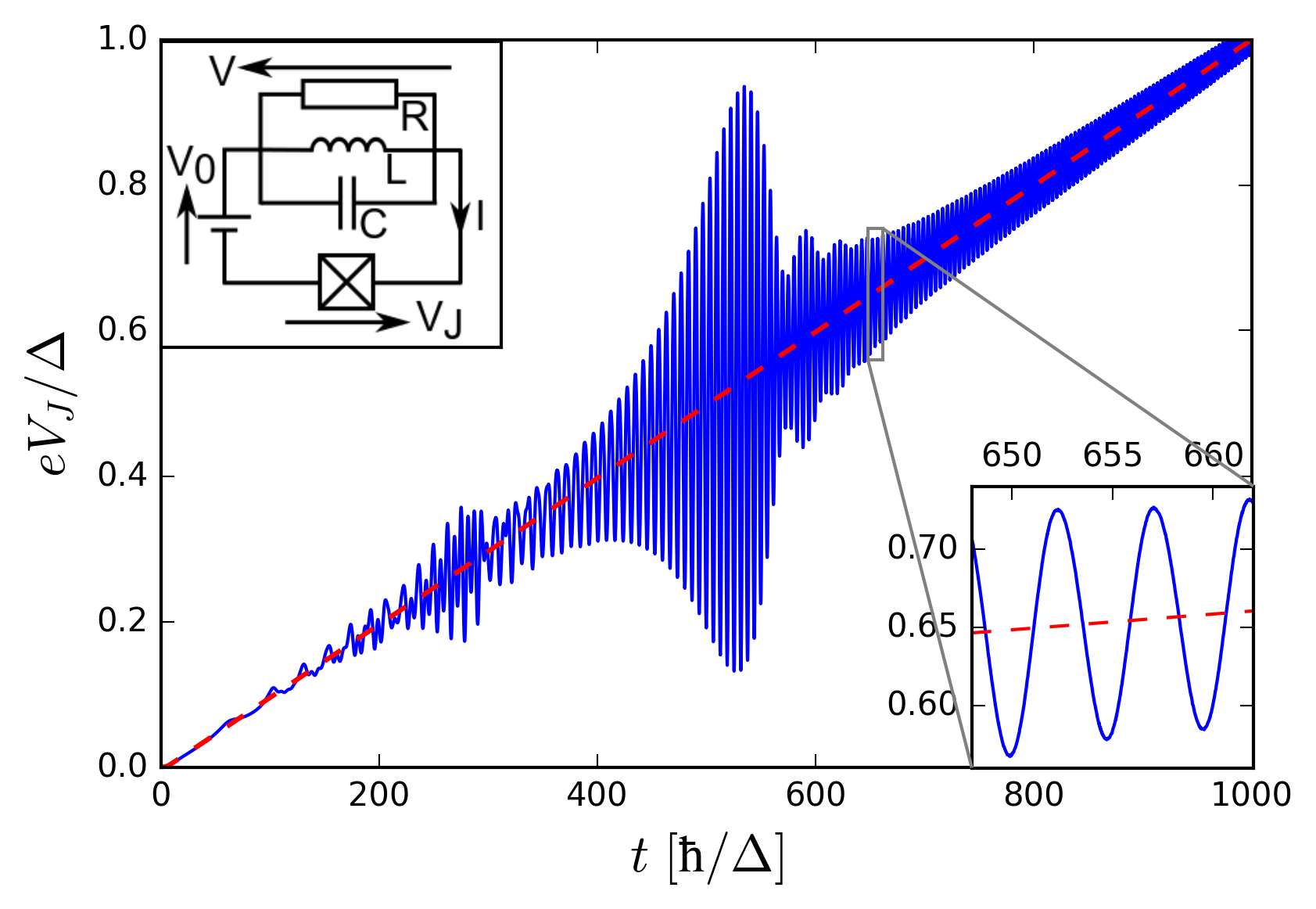}
\vspace{-3ex}
\caption{RLC-BdG model. Upper inset: schematic of the RLC circuit. Main panel: voltage $V_J(t)$ across the junction (blue line) versus time $t$ for a linear voltage ramp in $V_0(t)$ (red dashed line). $Q=20$, $\omega_0 = \Delta$ and $R=3h/2e^2 \simeq 38.7\,\text{k}\Omega$. Bottom inset: zoom of the main curve showing the ac Josephson effect oscillations. The resonance of the RLC circuit is visible for $eV_J=\hbar\omega_0/2$ and $eV_J=\hbar\omega_0/4$}
\label{fig3}
\end{figure}

We now turn to the full simulation of the RC-BdG model. The bare simulation data is shown in Fig. \ref{fig1} where the dashed line corresponds to a slow (quasi-static) ramp of $I_0$ so that the entire $I-V$ characteristics of the device can be extracted from a single simulation. We ramp the current first up and then down to zero in order to capture the hysteresis loop of the junction. The blue line corresponds to the voltage across the junction as a function of time. As shown in the inset, the blue line contains an important oscillating part that corresponds to the AC Josephson effect. From this data, we calculate the voltage $\bar V$ across the junction, averaged over a small time window (to get rid of the AC Josephson signal). Fig. \ref{fig2} show the resulting plot of $I_0$ versus $\bar V$ (blue line). The dotted line corresponds to the various asymptotic of the RCJ model discussed above while the dashed line corresponds to the pure BdG model in the absence of the electromagnetic environment. The pure BdG model displays the usual kinks characteristics of the opening of a new MAR channel \cite{Cuevas2002}. The RC-BdG simulations reconcile the two limits: the pure MAR curve at high bias and the supercurrent branch of the RCJ model at small bias. 
In the crossover between these two extreme limits, it provides the minimum model that captures all important physical contributions, hence quantitatively predicts the full hysteresis loop including the retrapping current. 
The most interesting features of the system show up in its dynamics. Recording the phase difference $\varphi (t)$ across the junction and the current $I(t)$ that flows through it, the dynamics is properly captured by the corresponding out-of-equilibrium current phase $I-\varphi$ relation obtained from the corresponding parametric plot. the result is shown in the upper panel of Fig. \ref{fig2}. Such out-of-equilibrium $I-\varphi$ could be reconstructed from a high-frequency measurement of the different harmonic of $V(t)$. As a reference,  Fig. \ref{fig2} includes the equilibrium $I-\varphi$ characteristics
of the junction (dotted line) obtained by taking all contributions into account (i.e. both Andreev bound states and the small contribution from the continuous part of the spectrum). This equilibrium $I-\varphi$ relation contains small deviations to the sinusoidal form. However, out-of-equilibrium relations can be strongly different from the simple sinusoidal shape. This is true in particular in the returning part of the hysteresis loop (red line, square, visible component of the second harmonic) and close
to the MAR cusps (yellow line, triangle, strongly non-sinusoidal). In these regimes, the excursions in voltage across the junction are wide (as can be seen directly from Fig. \ref{fig1}) and the junction effectively highly non-linear.

{\it Results for the RLC-BdG model.} We now turn to a second circuit where the junction is put in series with a classical $RLC$
resonator as sketched in the inset of Fig. \ref{fig3}. The electromagnetic circuit is slightly more complex than the previous
$RC$ model, but in return, the highly non-linear behavior shown in the previous example manifests itself already on DC observables.
The resonator has a quality factor $Q=R\sqrt{C/L}$ and a resonance pulsation $\omega_0= 1/\sqrt{LC}$. The corresponding impedance $Z(\omega)$ takes the form $R/Z(\omega) = 1+iQ [\omega/\omega_0-\omega_0/\omega]$ and filters frequency around $\omega_0$. Such an
environment has been studied in a series of recent experiments using tunnel junctions \cite{Westig2017, Simmonds2004, Reagor2016, Zaretskey2013}.
The RLC circuit provides a direct probe of the AC signal present in the system. We expect a main resonance for $2e V_J/\hbar = \omega_0$
when the AC Josephson effect drives the RLC circuit. Due to the non-linear character of the junction, higher harmonic of the AC
Josephson effects are generated, so that additional features are expected at $2e V_J/\hbar = \omega_0/n$. Likewise, the non linearities imply that the RLC circuit can also be driven parametrically at $2\omega_0$ leading to features at $e V_J/\hbar = \omega_0/n$.

\begin{figure}
\centering
\includegraphics[scale=0.47]{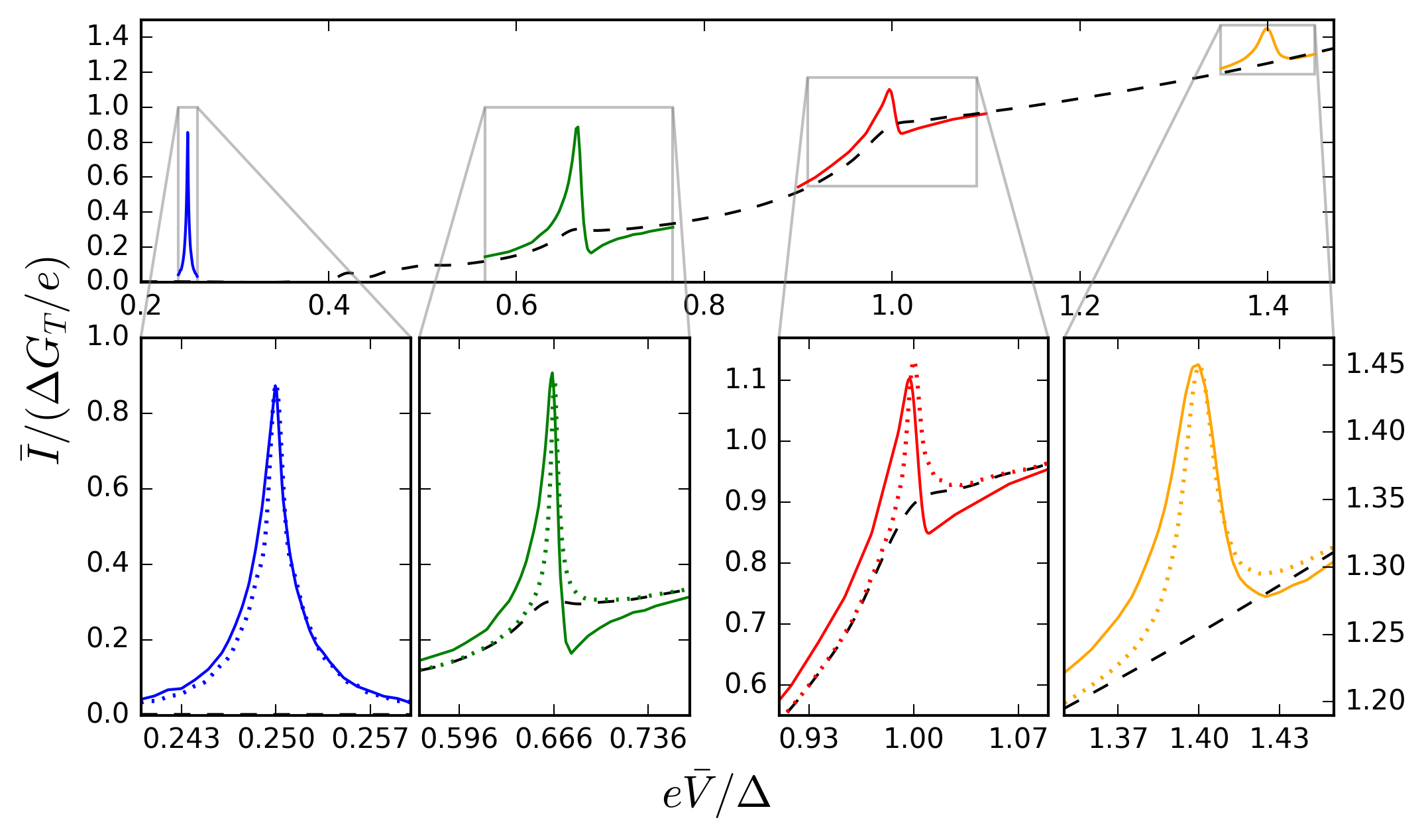}
\vspace{-3ex}
\caption{Current-voltage relation for four different resonator frequencies $\omega_0/\Delta = 1/4,2/3,1$ and $1.4$. Dashed line: $I_{\rm MAR}(V)$ in the absence of environment, thin color lines: RLC-BdG simulations, dotted lines: improved RLCJ model. Bottom panels: zoom of the main figure.}
\label{fig4}
\end{figure}

We compare the RLC-BdG calculations with an improved RLCJ model. The improved RLCJ model captures the super-current branch and the MAR non-linear I-V curve as,
\begin{equation}
I(\varphi) = I_c \sin\varphi + I_{\rm MAR}(2e\dot\varphi/\hbar)
\end{equation}
where $I_{\rm MAR}(V)$ is the DC non-linear $I-V$ characteristic of the junction in the absence of electromagnetic
environment (dashed line of Fig. \ref{fig4}). The numerical results for the average current $\bar I$ versus voltage
are shown in Fig. \ref{fig4} for four different RLC circuits with different frequencies $\omega_0$. We concentrate
on the main features around $2e V_J/\hbar = \omega_0$ and disregard the smaller peaks associated with higher harmonics and/or parametric pumping. The improved RLCJ model (dotted line) presents a Lorentzian like resonance at $2e V_J/\hbar = \omega_0$
for all four RLC circuits. When the resonance lies in the tunneling regime of the junction (blue line), there is a very good agreement between the improved RLCJ model and the full RLC-BdG simulations. The agreement is also qualitatively (but not quantitatively) good when the resonance corresponds to high voltages in the almost ``Ohmic'' regime of the junction (yellow line). However, for the two circuits where the resonance lies in the vicinity of a kink of the $I_{\rm MAR}(V)$ characteristic, the two models are strikingly different and the improved RLCJ model no longer applicable (green and red lines): the improved RLCJ model is typically off by $\pm 50\%$ including in the linewidth. 
In these regimes, we find that for 
$2e V_J/\hbar > \omega_0$, the current is
{\it reduced} with respect to $I_{\rm MAR}(V)$ instead of the Lorentzian increase observed in the improved RLCJ model. 
This reduction of the current is a direct manifestation of the non-linear AC physics happening in the device. This DC prediction is
the counterpart of the highly non-sinusoidal non-equilibrium current-phase relations discussed above for the RC-BdG case. However, the fact that the observable is in DC makes this prediction more easily accessible to an experimental test.

{\it Conclusion.} The {\it Environment-BdG} model presented in this manuscript unifies simple RCJ like models with microscopic models that include the quasi-particle spectrum of the junctions as well as its dynamics out-of-equilibrium. We have shown that the interplay between the two physics strongly modifies the behavior of the system and lead to new phenomena such as the voltage induced non-sinusoidal current phase relations.
Our approach provides a practical route to study the engineering of electromagnetic environments in the presence of junctions that go
beyond simple tunneling devices. Besides the example studied in this letter (a single channel junction with arbitrary transparency), other systems such as Josephson FETs \cite{Akazaki1996}, Majorana devices \cite{Sarma2015}, multi-terminal devices \cite{McCaughan2014} that are being developed by the community could be studied with the same technique \cite{Weston2015}. On the technical level, the approach could be extended to include electron-electron interaction and/or a self-consistent calculation of the superconducting gap at the time-dependent mean-field level \cite{Kloss2018}.

{\itshape Acknowledgment.} This work was supported by the ANR Fully Quantum (ANR-16-CE30-0015-02), ANR QTERA (ANR-15-CE24-0007-02), the NSF-ANR partnership PIRE: HYBRID and U.S. Office of Naval Research. We thank Manuel Houzet and Julia Meyer for useful discussions.

\bibliographystyle{apsrev4-1_own}
\bibliography{biblio}

\begin{thebibliography}{36}%
\makeatletter
\providecommand \@ifxundefined [1]{%
 \@ifx{#1\undefined}
}%
\providecommand \@ifnum [1]{%
 \ifnum #1\expandafter \@firstoftwo
 \else \expandafter \@secondoftwo
 \fi
}%
\providecommand \@ifx [1]{%
 \ifx #1\expandafter \@firstoftwo
 \else \expandafter \@secondoftwo
 \fi
}%
\providecommand \natexlab [1]{#1}%
\providecommand \enquote  [1]{``#1''}%
\providecommand \bibnamefont  [1]{#1}%
\providecommand \bibfnamefont [1]{#1}%
\providecommand \citenamefont [1]{#1}%
\providecommand \href@noop [0]{\@secondoftwo}%
\providecommand \href [0]{\begingroup \@sanitize@url \@href}%
\providecommand \@href[1]{\@@startlink{#1}\@@href}%
\providecommand \@@href[1]{\endgroup#1\@@endlink}%
\providecommand \@sanitize@url [0]{\catcode `\\12\catcode `\$12\catcode
  `\&12\catcode `\#12\catcode `\^12\catcode `\_12\catcode `\%12\relax}%
\providecommand \@@startlink[1]{}%
\providecommand \@@endlink[0]{}%
\providecommand \url  [0]{\begingroup\@sanitize@url \@url }%
\providecommand \@url [1]{\endgroup\@href {#1}{\urlprefix }}%
\providecommand \urlprefix  [0]{URL }%
\providecommand \Eprint [0]{\href }%
\providecommand \doibase [0]{http://dx.doi.org/}%
\providecommand \selectlanguage [0]{\@gobble}%
\providecommand \bibinfo  [0]{\@secondoftwo}%
\providecommand \bibfield  [0]{\@secondoftwo}%
\providecommand \translation [1]{[#1]}%
\providecommand \BibitemOpen [0]{}%
\providecommand \bibitemStop [0]{}%
\providecommand \bibitemNoStop [0]{.\EOS\space}%
\providecommand \EOS [0]{\spacefactor3000\relax}%
\providecommand \BibitemShut  [1]{\csname bibitem#1\endcsname}%
\let\auto@bib@innerbib\@empty
%</preamble>
\bibitem [{\citenamefont {Koch}\ \emph {et~al.}(2007)\citenamefont {Koch},
  \citenamefont {Yu}, \citenamefont {Gambetta}, \citenamefont {Houck},
  \citenamefont {Schuster}, \citenamefont {Majer}, \citenamefont {Blais},
  \citenamefont {Devoret}, \citenamefont {Girvin},\ and\ \citenamefont
  {Schoelkopf}}]{Koch2007}%
  \BibitemOpen
  \bibfield  {author} {\bibinfo {author} {\bibfnamefont {J.}~\bibnamefont
  {Koch}}, \bibinfo {author} {\bibfnamefont {T.~M.}\ \bibnamefont {Yu}},
  \bibinfo {author} {\bibfnamefont {J.}~\bibnamefont {Gambetta}}, \bibinfo
  {author} {\bibfnamefont {A.~A.}\ \bibnamefont {Houck}}, \bibinfo {author}
  {\bibfnamefont {D.~I.}\ \bibnamefont {Schuster}}, \bibinfo {author}
  {\bibfnamefont {J.}~\bibnamefont {Majer}}, \bibinfo {author} {\bibfnamefont
  {A.}~\bibnamefont {Blais}}, \bibinfo {author} {\bibfnamefont {M.~H.}\
  \bibnamefont {Devoret}}, \bibinfo {author} {\bibfnamefont {S.~M.}\
  \bibnamefont {Girvin}}, \ and\ \bibinfo {author} {\bibfnamefont {R.~J.}\
  \bibnamefont {Schoelkopf}},\ }\bibfield  {title} {\emph {\enquote {\bibinfo
  {title} {Charge-insensitive qubit design derived from the cooper pair box},}\
  }}\href {\doibase 10.1103/physreva.76.042319} {\bibfield  {journal} {\bibinfo
   {journal} {Physical Review A}\ }\textbf {\bibinfo {volume} {76}},\ \bibinfo
  {pages} {042319} (\bibinfo {year} {2007})}\BibitemShut {NoStop}%
\bibitem [{\citenamefont {Barends}\ \emph {et~al.}(2013)\citenamefont
  {Barends}, \citenamefont {Kelly}, \citenamefont {Megrant}, \citenamefont
  {Sank}, \citenamefont {Jeffrey}, \citenamefont {Chen}, \citenamefont {Yin},
  \citenamefont {Chiaro}, \citenamefont {Mutus}, \citenamefont {Neill},
  \citenamefont {O'Malley}, \citenamefont {Roushan}, \citenamefont {Wenner},
  \citenamefont {White}, \citenamefont {Cleland},\ and\ \citenamefont
  {Martinis}}]{Barends2013}%
  \BibitemOpen
  \bibfield  {author} {\bibinfo {author} {\bibfnamefont {R.}~\bibnamefont
  {Barends}}, \bibinfo {author} {\bibfnamefont {J.}~\bibnamefont {Kelly}},
  \bibinfo {author} {\bibfnamefont {A.}~\bibnamefont {Megrant}}, \bibinfo
  {author} {\bibfnamefont {D.}~\bibnamefont {Sank}}, \bibinfo {author}
  {\bibfnamefont {E.}~\bibnamefont {Jeffrey}}, \bibinfo {author} {\bibfnamefont
  {Y.}~\bibnamefont {Chen}}, \bibinfo {author} {\bibfnamefont {Y.}~\bibnamefont
  {Yin}}, \bibinfo {author} {\bibfnamefont {B.}~\bibnamefont {Chiaro}},
  \bibinfo {author} {\bibfnamefont {J.}~\bibnamefont {Mutus}}, \bibinfo
  {author} {\bibfnamefont {C.}~\bibnamefont {Neill}}, \bibinfo {author}
  {\bibfnamefont {P.}~\bibnamefont {O'Malley}}, \bibinfo {author}
  {\bibfnamefont {P.}~\bibnamefont {Roushan}}, \bibinfo {author} {\bibfnamefont
  {J.}~\bibnamefont {Wenner}}, \bibinfo {author} {\bibfnamefont {T.~C.}\
  \bibnamefont {White}}, \bibinfo {author} {\bibfnamefont {A.~N.}\ \bibnamefont
  {Cleland}}, \ and\ \bibinfo {author} {\bibfnamefont {J.~M.}\ \bibnamefont
  {Martinis}},\ }\bibfield  {title} {\emph {\enquote {\bibinfo {title}
  {Coherent josephson qubit suitable for scalable quantum integrated
  circuits},}\ }}\href {\doibase 10.1103/physrevlett.111.080502} {\bibfield
  {journal} {\bibinfo  {journal} {Physical Review Letters}\ }\textbf {\bibinfo
  {volume} {111}},\ \bibinfo {pages} {080502} (\bibinfo {year}
  {2013})}\BibitemShut {NoStop}%
\bibitem [{\citenamefont {Paik}\ \emph {et~al.}(2011)\citenamefont {Paik},
  \citenamefont {Schuster}, \citenamefont {Bishop}, \citenamefont {Kirchmair},
  \citenamefont {Catelani}, \citenamefont {Sears}, \citenamefont {Johnson},
  \citenamefont {Reagor}, \citenamefont {Frunzio}, \citenamefont {Glazman},
  \citenamefont {Girvin}, \citenamefont {Devoret},\ and\ \citenamefont
  {Schoelkopf}}]{Paik2011}%
  \BibitemOpen
  \bibfield  {author} {\bibinfo {author} {\bibfnamefont {H.}~\bibnamefont
  {Paik}}, \bibinfo {author} {\bibfnamefont {D.~I.}\ \bibnamefont {Schuster}},
  \bibinfo {author} {\bibfnamefont {L.~S.}\ \bibnamefont {Bishop}}, \bibinfo
  {author} {\bibfnamefont {G.}~\bibnamefont {Kirchmair}}, \bibinfo {author}
  {\bibfnamefont {G.}~\bibnamefont {Catelani}}, \bibinfo {author}
  {\bibfnamefont {A.~P.}\ \bibnamefont {Sears}}, \bibinfo {author}
  {\bibfnamefont {B.~R.}\ \bibnamefont {Johnson}}, \bibinfo {author}
  {\bibfnamefont {M.~J.}\ \bibnamefont {Reagor}}, \bibinfo {author}
  {\bibfnamefont {L.}~\bibnamefont {Frunzio}}, \bibinfo {author} {\bibfnamefont
  {L.~I.}\ \bibnamefont {Glazman}}, \bibinfo {author} {\bibfnamefont {S.~M.}\
  \bibnamefont {Girvin}}, \bibinfo {author} {\bibfnamefont {M.~H.}\
  \bibnamefont {Devoret}}, \ and\ \bibinfo {author} {\bibfnamefont {R.~J.}\
  \bibnamefont {Schoelkopf}},\ }\bibfield  {title} {\emph {\enquote {\bibinfo
  {title} {Observation of high coherence in josephson junction qubits measured
  in a three-dimensional circuit {QED} architecture},}\ }}\href {\doibase
  10.1103/physrevlett.107.240501} {\bibfield  {journal} {\bibinfo  {journal}
  {Physical Review Letters}\ }\textbf {\bibinfo {volume} {107}},\ \bibinfo
  {pages} {240501} (\bibinfo {year} {2011})}\BibitemShut {NoStop}%
\bibitem [{\citenamefont {Goffman}\ \emph {et~al.}(2000)\citenamefont
  {Goffman}, \citenamefont {Cron}, \citenamefont {Yeyati}, \citenamefont
  {Joyez}, \citenamefont {Devoret}, \citenamefont {Esteve},\ and\ \citenamefont
  {Urbina}}]{Goffman2000}%
  \BibitemOpen
  \bibfield  {author} {\bibinfo {author} {\bibfnamefont {M.~F.}\ \bibnamefont
  {Goffman}}, \bibinfo {author} {\bibfnamefont {R.}~\bibnamefont {Cron}},
  \bibinfo {author} {\bibfnamefont {A.~L.}\ \bibnamefont {Yeyati}}, \bibinfo
  {author} {\bibfnamefont {P.}~\bibnamefont {Joyez}}, \bibinfo {author}
  {\bibfnamefont {M.~H.}\ \bibnamefont {Devoret}}, \bibinfo {author}
  {\bibfnamefont {D.}~\bibnamefont {Esteve}}, \ and\ \bibinfo {author}
  {\bibfnamefont {C.}~\bibnamefont {Urbina}},\ }\bibfield  {title} {\emph
  {\enquote {\bibinfo {title} {Supercurrent in atomic point contacts and
  andreev states},}\ }}\href {\doibase 10.1103/physrevlett.85.170} {\bibfield
  {journal} {\bibinfo  {journal} {Physical Review Letters}\ }\textbf {\bibinfo
  {volume} {85}},\ \bibinfo {pages} {170} (\bibinfo {year} {2000})}\BibitemShut
  {NoStop}%
\bibitem [{\citenamefont {Mourik}\ \emph {et~al.}(2012)\citenamefont {Mourik},
  \citenamefont {Zuo}, \citenamefont {Frolov}, \citenamefont {Plissard},
  \citenamefont {Bakkers},\ and\ \citenamefont {Kouwenhoven}}]{Mourik2012}%
  \BibitemOpen
  \bibfield  {author} {\bibinfo {author} {\bibfnamefont {V.}~\bibnamefont
  {Mourik}}, \bibinfo {author} {\bibfnamefont {K.}~\bibnamefont {Zuo}},
  \bibinfo {author} {\bibfnamefont {S.~M.}\ \bibnamefont {Frolov}}, \bibinfo
  {author} {\bibfnamefont {S.~R.}\ \bibnamefont {Plissard}}, \bibinfo {author}
  {\bibfnamefont {E.~P. A.~M.}\ \bibnamefont {Bakkers}}, \ and\ \bibinfo
  {author} {\bibfnamefont {L.~P.}\ \bibnamefont {Kouwenhoven}},\ }\bibfield
  {title} {\emph {\enquote {\bibinfo {title} {Signatures of majorana fermions
  in hybrid superconductor-semiconductor nanowire devices},}\ }}\href {\doibase
  10.1126/science.1222360} {\bibfield  {journal} {\bibinfo  {journal}
  {Science}\ }\textbf {\bibinfo {volume} {336}},\ \bibinfo {pages} {1003}
  (\bibinfo {year} {2012})}\BibitemShut {NoStop}%
\bibitem [{\citenamefont {Izyumov}\ \emph {et~al.}(2002)\citenamefont
  {Izyumov}, \citenamefont {Proshin},\ and\ \citenamefont
  {Khusainov}}]{Izyumov2002}%
  \BibitemOpen
  \bibfield  {author} {\bibinfo {author} {\bibfnamefont {Y.~A.}\ \bibnamefont
  {Izyumov}}, \bibinfo {author} {\bibfnamefont {Y.~N.}\ \bibnamefont
  {Proshin}}, \ and\ \bibinfo {author} {\bibfnamefont {M.~G.}\ \bibnamefont
  {Khusainov}},\ }\bibfield  {title} {\emph {\enquote {\bibinfo {title}
  {Competition between superconductivity and magnetism in
  ferromagnet-superconductor heterostructures},}\ }}\href {\doibase
  10.3367/ufnr.0172.200202a.0113} {\bibfield  {journal} {\bibinfo  {journal}
  {Uspekhi Fizicheskih Nauk}\ }\textbf {\bibinfo {volume} {172}},\ \bibinfo
  {pages} {113} (\bibinfo {year} {2002})}\BibitemShut {NoStop}%
\bibitem [{\citenamefont {Buzdin}(2005)}]{Buzdin2005}%
  \BibitemOpen
  \bibfield  {author} {\bibinfo {author} {\bibfnamefont {A.~I.}\ \bibnamefont
  {Buzdin}},\ }\bibfield  {title} {\emph {\enquote {\bibinfo {title} {Proximity
  effects in superconductor-ferromagnet heterostructures},}\ }}\href {\doibase
  10.1103/revmodphys.77.935} {\bibfield  {journal} {\bibinfo  {journal}
  {Reviews of Modern Physics}\ }\textbf {\bibinfo {volume} {77}},\ \bibinfo
  {pages} {935} (\bibinfo {year} {2005})}\BibitemShut {NoStop}%
\bibitem [{\citenamefont {Riwar}\ \emph {et~al.}(2016)\citenamefont {Riwar},
  \citenamefont {Houzet}, \citenamefont {Meyer},\ and\ \citenamefont
  {Nazarov}}]{Riwar2016}%
  \BibitemOpen
  \bibfield  {author} {\bibinfo {author} {\bibfnamefont {R.-P.}\ \bibnamefont
  {Riwar}}, \bibinfo {author} {\bibfnamefont {M.}~\bibnamefont {Houzet}},
  \bibinfo {author} {\bibfnamefont {J.~S.}\ \bibnamefont {Meyer}}, \ and\
  \bibinfo {author} {\bibfnamefont {Y.~V.}\ \bibnamefont {Nazarov}},\
  }\bibfield  {title} {\emph {\enquote {\bibinfo {title} {Multi-terminal
  josephson junctions as topological matter},}\ }}\href {\doibase
  10.1038/ncomms11167} {\bibfield  {journal} {\bibinfo  {journal} {Nature
  Communications}\ }\textbf {\bibinfo {volume} {7}},\ \bibinfo {pages} {11167}
  (\bibinfo {year} {2016})}\BibitemShut {NoStop}%
\bibitem [{\citenamefont {Golubov}\ \emph {et~al.}(2004)\citenamefont
  {Golubov}, \citenamefont {Kupriyanov},\ and\ \citenamefont
  {Il'ichev}}]{Golubov2004}%
  \BibitemOpen
  \bibfield  {author} {\bibinfo {author} {\bibfnamefont {A.~A.}\ \bibnamefont
  {Golubov}}, \bibinfo {author} {\bibfnamefont {M.~Y.}\ \bibnamefont
  {Kupriyanov}}, \ and\ \bibinfo {author} {\bibfnamefont {E.}~\bibnamefont
  {Il'ichev}},\ }\bibfield  {title} {\emph {\enquote {\bibinfo {title} {The
  current-phase relation in josephson junctions},}\ }}\href {\doibase
  10.1103/revmodphys.76.411} {\bibfield  {journal} {\bibinfo  {journal}
  {Reviews of Modern Physics}\ }\textbf {\bibinfo {volume} {76}},\ \bibinfo
  {pages} {411} (\bibinfo {year} {2004})}\BibitemShut {NoStop}%
\bibitem [{\citenamefont {Likharev}(1979)}]{Likharev1979}%
  \BibitemOpen
  \bibfield  {author} {\bibinfo {author} {\bibfnamefont {K.~K.}\ \bibnamefont
  {Likharev}},\ }\bibfield  {title} {\emph {\enquote {\bibinfo {title}
  {Superconducting weak links},}\ }}\href {\doibase 10.1103/revmodphys.51.101}
  {\bibfield  {journal} {\bibinfo  {journal} {Reviews of Modern Physics}\
  }\textbf {\bibinfo {volume} {51}},\ \bibinfo {pages} {101} (\bibinfo {year}
  {1979})}\BibitemShut {NoStop}%
\bibitem [{\citenamefont {Stewart}(1968)}]{Stewart1968}%
  \BibitemOpen
  \bibfield  {author} {\bibinfo {author} {\bibfnamefont {W.~C.}\ \bibnamefont
  {Stewart}},\ }\bibfield  {title} {\emph {\enquote {\bibinfo {title}
  {Current-voltage characteristics of josephson junctions},}\ }}\href {\doibase
  10.1063/1.1651991} {\bibfield  {journal} {\bibinfo  {journal} {Applied
  Physics Letters}\ }\textbf {\bibinfo {volume} {12}},\ \bibinfo {pages} {277}
  (\bibinfo {year} {1968})}\BibitemShut {NoStop}%
\bibitem [{\citenamefont {Ben-Jacob}\ \emph {et~al.}(1982)\citenamefont
  {Ben-Jacob}, \citenamefont {Bergman}, \citenamefont {Matkowsky},\ and\
  \citenamefont {Schuss}}]{BenJacob1982}%
  \BibitemOpen
  \bibfield  {author} {\bibinfo {author} {\bibfnamefont {E.}~\bibnamefont
  {Ben-Jacob}}, \bibinfo {author} {\bibfnamefont {D.~J.}\ \bibnamefont
  {Bergman}}, \bibinfo {author} {\bibfnamefont {B.~J.}\ \bibnamefont
  {Matkowsky}}, \ and\ \bibinfo {author} {\bibfnamefont {Z.}~\bibnamefont
  {Schuss}},\ }\bibfield  {title} {\emph {\enquote {\bibinfo {title} {Lifetime
  of oscillatory steady states},}\ }}\href {\doibase 10.1103/physreva.26.2805}
  {\bibfield  {journal} {\bibinfo  {journal} {Physical Review A}\ }\textbf
  {\bibinfo {volume} {26}},\ \bibinfo {pages} {2805} (\bibinfo {year}
  {1982})}\BibitemShut {NoStop}%
\bibitem [{\citenamefont {Cassidy}\ \emph {et~al.}(2017)\citenamefont
  {Cassidy}, \citenamefont {Bruno}, \citenamefont {Rubbert}, \citenamefont
  {Irfan}, \citenamefont {Kammhuber}, \citenamefont {Schouten}, \citenamefont
  {Akhmerov},\ and\ \citenamefont {Kouwenhoven}}]{Cassidy2017}%
  \BibitemOpen
  \bibfield  {author} {\bibinfo {author} {\bibfnamefont {M.~C.}\ \bibnamefont
  {Cassidy}}, \bibinfo {author} {\bibfnamefont {A.}~\bibnamefont {Bruno}},
  \bibinfo {author} {\bibfnamefont {S.}~\bibnamefont {Rubbert}}, \bibinfo
  {author} {\bibfnamefont {M.}~\bibnamefont {Irfan}}, \bibinfo {author}
  {\bibfnamefont {J.}~\bibnamefont {Kammhuber}}, \bibinfo {author}
  {\bibfnamefont {R.~N.}\ \bibnamefont {Schouten}}, \bibinfo {author}
  {\bibfnamefont {A.~R.}\ \bibnamefont {Akhmerov}}, \ and\ \bibinfo {author}
  {\bibfnamefont {L.~P.}\ \bibnamefont {Kouwenhoven}},\ }\bibfield  {title}
  {\emph {\enquote {\bibinfo {title} {Demonstration of an ac josephson junction
  laser},}\ }}\href {\doibase 10.1126/science.aah6640} {\bibfield  {journal}
  {\bibinfo  {journal} {Science}\ }\textbf {\bibinfo {volume} {355}},\ \bibinfo
  {pages} {939} (\bibinfo {year} {2017})}\BibitemShut {NoStop}%
\bibitem [{\citenamefont {Xiang}\ \emph {et~al.}(2013)\citenamefont {Xiang},
  \citenamefont {Ashhab}, \citenamefont {You},\ and\ \citenamefont
  {Nori}}]{Xiang2013}%
  \BibitemOpen
  \bibfield  {author} {\bibinfo {author} {\bibfnamefont {Z.-L.}\ \bibnamefont
  {Xiang}}, \bibinfo {author} {\bibfnamefont {S.}~\bibnamefont {Ashhab}},
  \bibinfo {author} {\bibfnamefont {J.~Q.}\ \bibnamefont {You}}, \ and\
  \bibinfo {author} {\bibfnamefont {F.}~\bibnamefont {Nori}},\ }\bibfield
  {title} {\emph {\enquote {\bibinfo {title} {Hybrid quantum circuits:
  Superconducting circuits interacting with other quantum systems},}\ }}\href
  {\doibase 10.1103/revmodphys.85.623} {\bibfield  {journal} {\bibinfo
  {journal} {Reviews of Modern Physics}\ }\textbf {\bibinfo {volume} {85}},\
  \bibinfo {pages} {623} (\bibinfo {year} {2013})}\BibitemShut {NoStop}%
\bibitem [{\citenamefont {Devoret}\ \emph {et~al.}(1990)\citenamefont
  {Devoret}, \citenamefont {Esteve}, \citenamefont {Grabert}, \citenamefont
  {Ingold}, \citenamefont {Pothier},\ and\ \citenamefont
  {Urbina}}]{Devoret1990}%
  \BibitemOpen
  \bibfield  {author} {\bibinfo {author} {\bibfnamefont {M.~H.}\ \bibnamefont
  {Devoret}}, \bibinfo {author} {\bibfnamefont {D.}~\bibnamefont {Esteve}},
  \bibinfo {author} {\bibfnamefont {H.}~\bibnamefont {Grabert}}, \bibinfo
  {author} {\bibfnamefont {G.-L.}\ \bibnamefont {Ingold}}, \bibinfo {author}
  {\bibfnamefont {H.}~\bibnamefont {Pothier}}, \ and\ \bibinfo {author}
  {\bibfnamefont {C.}~\bibnamefont {Urbina}},\ }\bibfield  {title} {\emph
  {\enquote {\bibinfo {title} {Effect of the electromagnetic environment on the
  coulomb blockade in ultrasmall tunnel junctions},}\ }}\href {\doibase
  10.1103/physrevlett.64.1824} {\bibfield  {journal} {\bibinfo  {journal}
  {Physical Review Letters}\ }\textbf {\bibinfo {volume} {64}},\ \bibinfo
  {pages} {1824} (\bibinfo {year} {1990})}\BibitemShut {NoStop}%
\bibitem [{\citenamefont {Klapwijk}\ \emph {et~al.}(1982)\citenamefont
  {Klapwijk}, \citenamefont {Blonder},\ and\ \citenamefont
  {Tinkham}}]{Klapwijk1982}%
  \BibitemOpen
  \bibfield  {author} {\bibinfo {author} {\bibfnamefont {T.}~\bibnamefont
  {Klapwijk}}, \bibinfo {author} {\bibfnamefont {G.}~\bibnamefont {Blonder}}, \
  and\ \bibinfo {author} {\bibfnamefont {M.}~\bibnamefont {Tinkham}},\
  }\bibfield  {title} {\emph {\enquote {\bibinfo {title} {Explanation of
  subharmonic energy gap structure in superconducting contacts},}\ }}\href
  {\doibase 10.1016/0378-4363(82)90189-9} {\bibfield  {journal} {\bibinfo
  {journal} {Physica B$+$C}\ }\textbf {\bibinfo {volume} {109-110}},\ \bibinfo
  {pages} {1657} (\bibinfo {year} {1982})}\BibitemShut {NoStop}%
\bibitem [{\citenamefont {Averin}\ and\ \citenamefont
  {Bardas}(1995)}]{Averin1995}%
  \BibitemOpen
  \bibfield  {author} {\bibinfo {author} {\bibfnamefont {D.}~\bibnamefont
  {Averin}}\ and\ \bibinfo {author} {\bibfnamefont {A.}~\bibnamefont
  {Bardas}},\ }\bibfield  {title} {\emph {\enquote {\bibinfo {title} {ac
  josephson effect in a single quantum channel},}\ }}\href {\doibase
  10.1103/physrevlett.75.1831} {\bibfield  {journal} {\bibinfo  {journal}
  {Physical Review Letters}\ }\textbf {\bibinfo {volume} {75}},\ \bibinfo
  {pages} {1831} (\bibinfo {year} {1995})}\BibitemShut {NoStop}%
\bibitem [{\citenamefont {Cuevas}\ \emph {et~al.}(2002)\citenamefont {Cuevas},
  \citenamefont {Heurich}, \citenamefont {Mart{\'{\i}}n-Rodero}, \citenamefont
  {Yeyati},\ and\ \citenamefont {Sch\"{o}n}}]{Cuevas2002}%
  \BibitemOpen
  \bibfield  {author} {\bibinfo {author} {\bibfnamefont {J.~C.}\ \bibnamefont
  {Cuevas}}, \bibinfo {author} {\bibfnamefont {J.}~\bibnamefont {Heurich}},
  \bibinfo {author} {\bibfnamefont {A.}~\bibnamefont {Mart{\'{\i}}n-Rodero}},
  \bibinfo {author} {\bibfnamefont {A.~L.}\ \bibnamefont {Yeyati}}, \ and\
  \bibinfo {author} {\bibfnamefont {G.}~\bibnamefont {Sch\"{o}n}},\ }\bibfield
  {title} {\emph {\enquote {\bibinfo {title} {Subharmonic shapiro steps and
  assisted tunneling in superconducting point contacts},}\ }}\href {\doibase
  10.1103/physrevlett.88.157001} {\bibfield  {journal} {\bibinfo  {journal}
  {Physical Review Letters}\ }\textbf {\bibinfo {volume} {88}},\ \bibinfo
  {pages} {157001} (\bibinfo {year} {2002})}\BibitemShut {NoStop}%
\bibitem [{\citenamefont {Houzet}\ \emph {et~al.}(2013)\citenamefont {Houzet},
  \citenamefont {Meyer}, \citenamefont {Badiane},\ and\ \citenamefont
  {Glazman}}]{Houzet2013}%
  \BibitemOpen
  \bibfield  {author} {\bibinfo {author} {\bibfnamefont {M.}~\bibnamefont
  {Houzet}}, \bibinfo {author} {\bibfnamefont {J.~S.}\ \bibnamefont {Meyer}},
  \bibinfo {author} {\bibfnamefont {D.~M.}\ \bibnamefont {Badiane}}, \ and\
  \bibinfo {author} {\bibfnamefont {L.~I.}\ \bibnamefont {Glazman}},\
  }\bibfield  {title} {\emph {\enquote {\bibinfo {title} {Dynamics of majorana
  states in a topological josephson junction},}\ }}\href {\doibase
  10.1103/physrevlett.111.046401} {\bibfield  {journal} {\bibinfo  {journal}
  {Physical Review Letters}\ }\textbf {\bibinfo {volume} {111}},\ \bibinfo
  {pages} {046401} (\bibinfo {year} {2013})}\BibitemShut {NoStop}%
\bibitem [{\citenamefont {Weston}\ and\ \citenamefont
  {Waintal}(2016{\natexlab{a}})}]{Weston2016}%
  \BibitemOpen
  \bibfield  {author} {\bibinfo {author} {\bibfnamefont {J.}~\bibnamefont
  {Weston}}\ and\ \bibinfo {author} {\bibfnamefont {X.}~\bibnamefont
  {Waintal}},\ }\bibfield  {title} {\emph {\enquote {\bibinfo {title}
  {Linear-scaling source-sink algorithm for simulating time-resolved quantum
  transport and superconductivity},}\ }}\href {\doibase
  10.1103/physrevb.93.134506} {\bibfield  {journal} {\bibinfo  {journal}
  {Physical Review B}\ }\textbf {\bibinfo {volume} {93}},\ \bibinfo {pages}
  {134506} (\bibinfo {year} {2016}{\natexlab{a}})}\BibitemShut {NoStop}%
\bibitem [{\citenamefont {Kloss}\ \emph {et~al.}(2018)\citenamefont {Kloss},
  \citenamefont {Weston},\ and\ \citenamefont {Waintal}}]{Kloss2018}%
  \BibitemOpen
  \bibfield  {author} {\bibinfo {author} {\bibfnamefont {T.}~\bibnamefont
  {Kloss}}, \bibinfo {author} {\bibfnamefont {J.}~\bibnamefont {Weston}}, \
  and\ \bibinfo {author} {\bibfnamefont {X.}~\bibnamefont {Waintal}},\
  }\bibfield  {title} {\emph {\enquote {\bibinfo {title} {Transient and sharvin
  resistances of luttinger liquids},}\ }}\href {\doibase
  10.1103/physrevb.97.165134} {\bibfield  {journal} {\bibinfo  {journal}
  {Physical Review B}\ }\textbf {\bibinfo {volume} {97}},\ \bibinfo {pages}
  {165134} (\bibinfo {year} {2018})}\BibitemShut {NoStop}%
\bibitem [{\citenamefont {Gaury}\ and\ \citenamefont
  {Waintal}(2016)}]{Gaury2016}%
  \BibitemOpen
  \bibfield  {author} {\bibinfo {author} {\bibfnamefont {B.}~\bibnamefont
  {Gaury}}\ and\ \bibinfo {author} {\bibfnamefont {X.}~\bibnamefont
  {Waintal}},\ }\bibfield  {title} {\emph {\enquote {\bibinfo {title} {A
  computational approach to quantum noise in time-dependent nanoelectronic
  devices},}\ }}\href {\doibase 10.1016/j.physe.2015.09.009} {\bibfield
  {journal} {\bibinfo  {journal} {Physica E: Low-dimensional Systems and
  Nanostructures}\ }\textbf {\bibinfo {volume} {75}},\ \bibinfo {pages} {72}
  (\bibinfo {year} {2016})}\BibitemShut {NoStop}%
\bibitem [{\citenamefont {Gaury}\ \emph {et~al.}(2014)\citenamefont {Gaury},
  \citenamefont {Weston}, \citenamefont {Santin}, \citenamefont {Houzet},
  \citenamefont {Groth},\ and\ \citenamefont {Waintal}}]{Gaury2014}%
  \BibitemOpen
  \bibfield  {author} {\bibinfo {author} {\bibfnamefont {B.}~\bibnamefont
  {Gaury}}, \bibinfo {author} {\bibfnamefont {J.}~\bibnamefont {Weston}},
  \bibinfo {author} {\bibfnamefont {M.}~\bibnamefont {Santin}}, \bibinfo
  {author} {\bibfnamefont {M.}~\bibnamefont {Houzet}}, \bibinfo {author}
  {\bibfnamefont {C.}~\bibnamefont {Groth}}, \ and\ \bibinfo {author}
  {\bibfnamefont {X.}~\bibnamefont {Waintal}},\ }\bibfield  {title} {\emph
  {\enquote {\bibinfo {title} {Numerical simulations of time-resolved quantum
  electronics},}\ }}\href {\doibase 10.1016/j.physrep.2013.09.001} {\bibfield
  {journal} {\bibinfo  {journal} {Physics Reports}\ }\textbf {\bibinfo {volume}
  {534}},\ \bibinfo {pages} {1} (\bibinfo {year} {2014})}\BibitemShut {NoStop}%
\bibitem [{\citenamefont {Weston}\ and\ \citenamefont
  {Waintal}(2016{\natexlab{b}})}]{Weston2016b}%
  \BibitemOpen
  \bibfield  {author} {\bibinfo {author} {\bibfnamefont {J.}~\bibnamefont
  {Weston}}\ and\ \bibinfo {author} {\bibfnamefont {X.}~\bibnamefont
  {Waintal}},\ }\bibfield  {title} {\emph {\enquote {\bibinfo {title} {Towards
  realistic time-resolved simulations of quantum devices},}\ }}\href {\doibase
  10.1007/s10825-016-0855-9} {\bibfield  {journal} {\bibinfo  {journal}
  {Journal of Computational Electronics}\ }\textbf {\bibinfo {volume} {15}},\
  \bibinfo {pages} {1148} (\bibinfo {year} {2016}{\natexlab{b}})}\BibitemShut
  {NoStop}%
\bibitem [{\citenamefont {Groth}\ \emph {et~al.}(2014)\citenamefont {Groth},
  \citenamefont {Wimmer}, \citenamefont {Akhmerov},\ and\ \citenamefont
  {Waintal}}]{Groth2014}%
  \BibitemOpen
  \bibfield  {author} {\bibinfo {author} {\bibfnamefont {C.~W.}\ \bibnamefont
  {Groth}}, \bibinfo {author} {\bibfnamefont {M.}~\bibnamefont {Wimmer}},
  \bibinfo {author} {\bibfnamefont {A.~R.}\ \bibnamefont {Akhmerov}}, \ and\
  \bibinfo {author} {\bibfnamefont {X.}~\bibnamefont {Waintal}},\ }\bibfield
  {title} {\emph {\enquote {\bibinfo {title} {Kwant: a software package for
  quantum transport},}\ }}\href {\doibase 10.1088/1367-2630/16/6/063065}
  {\bibfield  {journal} {\bibinfo  {journal} {New Journal of Physics}\ }\textbf
  {\bibinfo {volume} {16}},\ \bibinfo {pages} {063065} (\bibinfo {year}
  {2014})}\BibitemShut {NoStop}%
\bibitem [{\citenamefont {Hairer}\ \emph {et~al.}(1993)\citenamefont {Hairer},
  \citenamefont {Wanner},\ and\ \citenamefont {Norsett}}]{Hairer1993}%
  \BibitemOpen
  \bibfield  {author} {\bibinfo {author} {\bibfnamefont {E.}~\bibnamefont
  {Hairer}}, \bibinfo {author} {\bibfnamefont {G.}~\bibnamefont {Wanner}}, \
  and\ \bibinfo {author} {\bibfnamefont {S.~P.}\ \bibnamefont {Norsett}},\
  }\href {\doibase 10.1007/978-3-540-78862-1} {\emph {\bibinfo {title} {Solving
  Ordinary Differential Equations I}}}\ (\bibinfo  {publisher} {Springer Berlin
  Heidelberg},\ \bibinfo {year} {1993})\BibitemShut {NoStop}%
\bibitem [{\citenamefont {Istas}\ \emph {et~al.}(2018)\citenamefont {Istas},
  \citenamefont {Groth}, \citenamefont {Akhmerov}, \citenamefont {Wimmer},\
  and\ \citenamefont {Waintal}}]{Istas2018}%
  \BibitemOpen
  \bibfield  {author} {\bibinfo {author} {\bibfnamefont {M.}~\bibnamefont
  {Istas}}, \bibinfo {author} {\bibfnamefont {C.}~\bibnamefont {Groth}},
  \bibinfo {author} {\bibfnamefont {A.}~\bibnamefont {Akhmerov}}, \bibinfo
  {author} {\bibfnamefont {M.}~\bibnamefont {Wimmer}}, \ and\ \bibinfo {author}
  {\bibfnamefont {X.}~\bibnamefont {Waintal}},\ }\bibfield  {title} {\emph
  {\enquote {\bibinfo {title} {A general algorithm for computing bound states
  in infinite tight-binding systems},}\ }}\href {\doibase
  10.21468/scipostphys.4.5.026} {\bibfield  {journal} {\bibinfo  {journal}
  {{SciPost} Physics}\ }\textbf {\bibinfo {volume} {4}},\ \bibinfo {pages}
  {026} (\bibinfo {year} {2018})}\BibitemShut {NoStop}%
\bibitem [{\citenamefont {Rossignol}\ \emph {et~al.}(2019)\citenamefont
  {Rossignol}, \citenamefont {Thomas}, \citenamefont {Weston}, \citenamefont
  {Gaury}, \citenamefont {Groth},\ and\ \citenamefont {Waintal}}]{code}%
  \BibitemOpen
  \bibfield  {author} {\bibinfo {author} {\bibfnamefont {B.}~\bibnamefont
  {Rossignol}}, \bibinfo {author} {\bibfnamefont {K.}~\bibnamefont {Thomas}},
  \bibinfo {author} {\bibfnamefont {J.}~\bibnamefont {Weston}}, \bibinfo
  {author} {\bibfnamefont {B.}~\bibnamefont {Gaury}}, \bibinfo {author}
  {\bibfnamefont {C.}~\bibnamefont {Groth}}, \ and\ \bibinfo {author}
  {\bibfnamefont {X.}~\bibnamefont {Waintal}},\ }\href {\doibase
  10.5281/zenodo.2543545} {\enquote {\bibinfo {title} {Role of the
  quasi-particles in an electric circuit with josephson junctions (code and
  data)},}\ } (\bibinfo {year} {2019})\BibitemShut {NoStop}%
\bibitem [{\citenamefont {Weston}\ \emph {et~al.}(2015)\citenamefont {Weston},
  \citenamefont {Gaury},\ and\ \citenamefont {Waintal}}]{Weston2015}%
  \BibitemOpen
  \bibfield  {author} {\bibinfo {author} {\bibfnamefont {J.}~\bibnamefont
  {Weston}}, \bibinfo {author} {\bibfnamefont {B.}~\bibnamefont {Gaury}}, \
  and\ \bibinfo {author} {\bibfnamefont {X.}~\bibnamefont {Waintal}},\
  }\bibfield  {title} {\emph {\enquote {\bibinfo {title} {Manipulating andreev
  and majorana bound states with microwaves},}\ }}\href {\doibase
  10.1103/physrevb.92.020513} {\bibfield  {journal} {\bibinfo  {journal}
  {Physical Review B}\ }\textbf {\bibinfo {volume} {92}},\ \bibinfo {pages}
  {020513} (\bibinfo {year} {2015})}\BibitemShut {NoStop}%
\bibitem [{\citenamefont {Westig}\ \emph {et~al.}(2017)\citenamefont {Westig},
  \citenamefont {Kubala}, \citenamefont {Parlavecchio}, \citenamefont
  {Mukharsky}, \citenamefont {Altimiras}, \citenamefont {Joyez}, \citenamefont
  {Vion}, \citenamefont {Roche}, \citenamefont {Esteve}, \citenamefont
  {Hofheinz}, \citenamefont {Trif}, \citenamefont {Simon}, \citenamefont
  {Ankerhold},\ and\ \citenamefont {Portier}}]{Westig2017}%
  \BibitemOpen
  \bibfield  {author} {\bibinfo {author} {\bibfnamefont {M.}~\bibnamefont
  {Westig}}, \bibinfo {author} {\bibfnamefont {B.}~\bibnamefont {Kubala}},
  \bibinfo {author} {\bibfnamefont {O.}~\bibnamefont {Parlavecchio}}, \bibinfo
  {author} {\bibfnamefont {Y.}~\bibnamefont {Mukharsky}}, \bibinfo {author}
  {\bibfnamefont {C.}~\bibnamefont {Altimiras}}, \bibinfo {author}
  {\bibfnamefont {P.}~\bibnamefont {Joyez}}, \bibinfo {author} {\bibfnamefont
  {D.}~\bibnamefont {Vion}}, \bibinfo {author} {\bibfnamefont {P.}~\bibnamefont
  {Roche}}, \bibinfo {author} {\bibfnamefont {D.}~\bibnamefont {Esteve}},
  \bibinfo {author} {\bibfnamefont {M.}~\bibnamefont {Hofheinz}}, \bibinfo
  {author} {\bibfnamefont {M.}~\bibnamefont {Trif}}, \bibinfo {author}
  {\bibfnamefont {P.}~\bibnamefont {Simon}}, \bibinfo {author} {\bibfnamefont
  {J.}~\bibnamefont {Ankerhold}}, \ and\ \bibinfo {author} {\bibfnamefont
  {F.}~\bibnamefont {Portier}},\ }\bibfield  {title} {\emph {\enquote {\bibinfo
  {title} {Emission of nonclassical radiation by inelastic cooper pair
  tunneling},}\ }}\href {\doibase 10.1103/physrevlett.119.137001} {\bibfield
  {journal} {\bibinfo  {journal} {Physical Review Letters}\ }\textbf {\bibinfo
  {volume} {119}},\ \bibinfo {pages} {137001} (\bibinfo {year}
  {2017})}\BibitemShut {NoStop}%
\bibitem [{\citenamefont {Simmonds}\ \emph {et~al.}(2004)\citenamefont
  {Simmonds}, \citenamefont {Lang}, \citenamefont {Hite}, \citenamefont {Nam},
  \citenamefont {Pappas},\ and\ \citenamefont {Martinis}}]{Simmonds2004}%
  \BibitemOpen
  \bibfield  {author} {\bibinfo {author} {\bibfnamefont {R.}~\bibnamefont
  {Simmonds}}, \bibinfo {author} {\bibfnamefont {K.}~\bibnamefont {Lang}},
  \bibinfo {author} {\bibfnamefont {D.}~\bibnamefont {Hite}}, \bibinfo {author}
  {\bibfnamefont {S.}~\bibnamefont {Nam}}, \bibinfo {author} {\bibfnamefont
  {D.}~\bibnamefont {Pappas}}, \ and\ \bibinfo {author} {\bibfnamefont
  {J.}~\bibnamefont {Martinis}},\ }\bibfield  {title} {\emph {\enquote
  {\bibinfo {title} {Decoherence in josephson phase qubits from junction
  resonators},}\ }}\href {\doibase 10.1103/physrevlett.93.077003} {\bibfield
  {journal} {\bibinfo  {journal} {Physical Review Letters}\ }\textbf {\bibinfo
  {volume} {93}},\ \bibinfo {pages} {077003} (\bibinfo {year}
  {2004})}\BibitemShut {NoStop}%
\bibitem [{\citenamefont {Reagor}\ \emph {et~al.}(2016)\citenamefont {Reagor},
  \citenamefont {Pfaff}, \citenamefont {Axline}, \citenamefont {Heeres},
  \citenamefont {Ofek}, \citenamefont {Sliwa}, \citenamefont {Holland},
  \citenamefont {Wang}, \citenamefont {Blumoff}, \citenamefont {Chou},
  \citenamefont {Hatridge}, \citenamefont {Frunzio}, \citenamefont {Devoret},
  \citenamefont {Jiang},\ and\ \citenamefont {Schoelkopf}}]{Reagor2016}%
  \BibitemOpen
  \bibfield  {author} {\bibinfo {author} {\bibfnamefont {M.}~\bibnamefont
  {Reagor}}, \bibinfo {author} {\bibfnamefont {W.}~\bibnamefont {Pfaff}},
  \bibinfo {author} {\bibfnamefont {C.}~\bibnamefont {Axline}}, \bibinfo
  {author} {\bibfnamefont {R.~W.}\ \bibnamefont {Heeres}}, \bibinfo {author}
  {\bibfnamefont {N.}~\bibnamefont {Ofek}}, \bibinfo {author} {\bibfnamefont
  {K.}~\bibnamefont {Sliwa}}, \bibinfo {author} {\bibfnamefont
  {E.}~\bibnamefont {Holland}}, \bibinfo {author} {\bibfnamefont
  {C.}~\bibnamefont {Wang}}, \bibinfo {author} {\bibfnamefont {J.}~\bibnamefont
  {Blumoff}}, \bibinfo {author} {\bibfnamefont {K.}~\bibnamefont {Chou}},
  \bibinfo {author} {\bibfnamefont {M.~J.}\ \bibnamefont {Hatridge}}, \bibinfo
  {author} {\bibfnamefont {L.}~\bibnamefont {Frunzio}}, \bibinfo {author}
  {\bibfnamefont {M.~H.}\ \bibnamefont {Devoret}}, \bibinfo {author}
  {\bibfnamefont {L.}~\bibnamefont {Jiang}}, \ and\ \bibinfo {author}
  {\bibfnamefont {R.~J.}\ \bibnamefont {Schoelkopf}},\ }\bibfield  {title}
  {\emph {\enquote {\bibinfo {title} {Quantum memory with millisecond coherence
  in circuit {QED}},}\ }}\href {\doibase 10.1103/physrevb.94.014506} {\bibfield
   {journal} {\bibinfo  {journal} {Physical Review B}\ }\textbf {\bibinfo
  {volume} {94}},\ \bibinfo {pages} {014506} (\bibinfo {year}
  {2016})}\BibitemShut {NoStop}%
\bibitem [{\citenamefont {Zaretskey}\ \emph {et~al.}(2013)\citenamefont
  {Zaretskey}, \citenamefont {Suri}, \citenamefont {Novikov}, \citenamefont
  {Wellstood},\ and\ \citenamefont {Palmer}}]{Zaretskey2013}%
  \BibitemOpen
  \bibfield  {author} {\bibinfo {author} {\bibfnamefont {V.}~\bibnamefont
  {Zaretskey}}, \bibinfo {author} {\bibfnamefont {B.}~\bibnamefont {Suri}},
  \bibinfo {author} {\bibfnamefont {S.}~\bibnamefont {Novikov}}, \bibinfo
  {author} {\bibfnamefont {F.~C.}\ \bibnamefont {Wellstood}}, \ and\ \bibinfo
  {author} {\bibfnamefont {B.~S.}\ \bibnamefont {Palmer}},\ }\bibfield  {title}
  {\emph {\enquote {\bibinfo {title} {Spectroscopy of a cooper-pair box coupled
  to a two-level system via charge and critical current},}\ }}\href {\doibase
  10.1103/physrevb.87.174522} {\bibfield  {journal} {\bibinfo  {journal}
  {Physical Review B}\ }\textbf {\bibinfo {volume} {87}},\ \bibinfo {pages}
  {174522} (\bibinfo {year} {2013})}\BibitemShut {NoStop}%
\bibitem [{\citenamefont {Akazaki}\ \emph {et~al.}(1996)\citenamefont
  {Akazaki}, \citenamefont {Takayanagi}, \citenamefont {Nitta},\ and\
  \citenamefont {Enoki}}]{Akazaki1996}%
  \BibitemOpen
  \bibfield  {author} {\bibinfo {author} {\bibfnamefont {T.}~\bibnamefont
  {Akazaki}}, \bibinfo {author} {\bibfnamefont {H.}~\bibnamefont {Takayanagi}},
  \bibinfo {author} {\bibfnamefont {J.}~\bibnamefont {Nitta}}, \ and\ \bibinfo
  {author} {\bibfnamefont {T.}~\bibnamefont {Enoki}},\ }\bibfield  {title}
  {\emph {\enquote {\bibinfo {title} {A josephson field effect transistor using
  an {InAs}-inserted-channel ${In_{0.52}Al_{0.48}As/In_{0.53}Ga_{0.47}As}$
  inverted modulation-doped structure},}\ }}\href {\doibase 10.1063/1.116704}
  {\bibfield  {journal} {\bibinfo  {journal} {Applied Physics Letters}\
  }\textbf {\bibinfo {volume} {68}},\ \bibinfo {pages} {418} (\bibinfo {year}
  {1996})}\BibitemShut {NoStop}%
\bibitem [{\citenamefont {Sarma}\ \emph {et~al.}(2015)\citenamefont {Sarma},
  \citenamefont {Freedman},\ and\ \citenamefont {Nayak}}]{Sarma2015}%
  \BibitemOpen
  \bibfield  {author} {\bibinfo {author} {\bibfnamefont {S.~D.}\ \bibnamefont
  {Sarma}}, \bibinfo {author} {\bibfnamefont {M.}~\bibnamefont {Freedman}}, \
  and\ \bibinfo {author} {\bibfnamefont {C.}~\bibnamefont {Nayak}},\ }\bibfield
   {title} {\emph {\enquote {\bibinfo {title} {Majorana zero modes and
  topological quantum computation},}\ }}\href {\doibase 10.1038/npjqi.2015.1}
  {\bibfield  {journal} {\bibinfo  {journal} {npj Quantum Information}\
  }\textbf {\bibinfo {volume} {1}},\ \bibinfo {pages} {15001} (\bibinfo {year}
  {2015})}\BibitemShut {NoStop}%
\bibitem [{\citenamefont {McCaughan}\ and\ \citenamefont
  {Berggren}(2014)}]{McCaughan2014}%
  \BibitemOpen
  \bibfield  {author} {\bibinfo {author} {\bibfnamefont {A.~N.}\ \bibnamefont
  {McCaughan}}\ and\ \bibinfo {author} {\bibfnamefont {K.~K.}\ \bibnamefont
  {Berggren}},\ }\bibfield  {title} {\emph {\enquote {\bibinfo {title} {A
  superconducting-nanowire three-terminal electrothermal device},}\ }}\href
  {\doibase 10.1021/nl502629x} {\bibfield  {journal} {\bibinfo  {journal} {Nano
  Letters}\ }\textbf {\bibinfo {volume} {14}},\ \bibinfo {pages} {5748}
  (\bibinfo {year} {2014})}\BibitemShut {NoStop}%
\end{thebibliography}%
\end{document}